# Analytical Gradient Theory for Resolvent-Fitted Second-Order Extended Multiconfiguration Perturbation Theory (XMCQDPT2)

Jae Woo Park[*]

*Department of Chemistry, Chungbuk National University (CBNU), Cheongju 28644, Korea*

**Abstract**

We present the formulation and implementation of an analytical gradient algorithm for extended multiconfiguration quasidegenerate perturbation theory (XMCQDPT2) with the resolvent-fitting approximation by Granovsky. This algorithm is powerful when optimizing molecular configurations with a moderate-sized active space and many electronic states. First, we present the powerfulness and accuracy of resolvent-fitting approximations compared to the canonical XMCQDPT2 theory. Then, we demonstrate the utility of the current algorithm in frequency analyses, optimizing the minimum energy conical intersection (MECI) geometries of the retinal chromophore model RPSB6, and evaluating nuclear gradients when there are many electronic states. Furthermore, we parallelize the algorithm using the OpenMP/MPI hybrid approach. Additionally, we report the computational cost and parallel efficiency of the program.

---

[*] E-mail: jaewoopark@cbnu.ac.kr

## 1. INTRODUCTION

Multireference perturbation theories (MRPTs) are an efficient means to recover dynamical correlations from multiconfiguration self-consistent field (MCSCF) reference functions. There are many MRPT variants: complete active space perturbation theory (CASPT),[1,2] *N*-electron valence state perturbation theory (NEVPT),[3-5] multireference Møller–Plesset perturbation theory (MRMP)[6-8] and its multistate extension, multiconfiguration quasidegenerate perturbation theory (MCQDPT)[6-9], which are among the most widely used.

The availability of analytical nuclear gradients expands the utility of quantum chemical methods, particularly in applications using geometry optimizations and molecular dynamics simulations.[10-13] Analytical gradient methods were developed for CASPT2,[14-22] NEVPT2,[23-26] and, most recently, RASPT2.[27] In fact, the first development of the analytical gradient for MRPTs occurred in MCQDPT2 by Nakano, Hirao, and Gordon.[28,29] Recently, we reported its applications in conical intersection optimizations of photochemical systems.[30]

MRPTs are classified into internally contracted and uncontracted methods. Internal contraction means that the first-order interacting space (FOIS) is formed by applying excitation operators onto the reference states. For example, CASPT2, SC- and PC-NEVPT2 are internally contracted, while UC-NEVPT2 and MCQDPT2 are uncontracted. The internally contracted methods are computationally more efficient, and the errors due to the internal contraction are in many cases marginal.[31-34] However, in some instances, such as when there are interactions between the states with small energetic gaps, the errors are significant.[35,36] In such cases, uncontracted theories can be useful; but only when it is practically possible in terms of computational cost.

The computational cost for evaluating the energy and nuclear gradient of MCQDPT2 is much higher than the cost of CASPT2 and NEVPT2, as MCQDPT2 is an uncontracted theory. The computational cost depends on the size of the active space $N_{act}$ and the number of electronic configurations $N_{det}$. The most expensive term dependent on $N_{det}$ in CASPT2 and NEVPT2 nuclear gradient evaluation is the four-particle reduced density matrix (4RDM), which scales as $N_{det}N_{act}^8$. In XMCQDPT2 nuclear gradient evaluation, on the other hand, the most expensive term scales as $N_{state}^2 N_{det} N_{act}^6 (N_{vir} + N_{closed})$.[30] In many CASCI problems, $N_{vir} + N_{closed} > N_{act}^2$, and treating systems with a moderate active space is usually impractical with MCQDPT2.

Two notable improvements over the original MCQDPT2 regarding the accuracy and computational efficiency were achieved by utilization of the "extension" and the resolvent fitting, respectively, both by Granovsky.[37,38] In the extended MCQDPT2 (XMCQDPT2), one uses the eigenstates of the Fock operator rather than the electronic Hamiltonian as the reference states, which results in the invariance of the perturbation theory with respect to the rotations among the states and thus the correct description of the potential energy surfaces near the surface crossing points. This concept is also applied to the MS-CASPT2 theory[39] to formulate XMS-CASPT2,[15] in which one can obtain the strict invariance with the so-called "MS-MR" internal contraction scheme.[18,24,40,41] The resolvent-fitting (or table-driven) approach decouples $N_{det}$ and $N_{vir}$ in the computational cost by introducing interpolation with $N_{grid}$ interpolation grids, reducing the computational cost of the most expensive term to $N_{grid}(N_{core} + N_{vir})N_{act}^6 + (N_{fgrid} + N_{state})N_{act}^6 N_{det} N_{state}$, where $N_{fgrid}$ grid points are included in the interpolation.[38] These contributions made (X)MCQDPT2 applicable to large systems with a large active space, a state-averaging (model) space, and basis functions. With analytical gradient, the resolvent-fitted XMCQDPT2 method can be a valuable alternative to the XMS-CASPT2 or QD-NEVPT2 theories, especially when strict invariance[24,40,41] is required.

In this work, we develop an analytical nuclear gradient theory for XMCQDPT2 with resolvent fitting. The analytical gradient is readily computed with moderate-sized active spaces up to (12$e$, 12$o$) with this approximation. As numerical examples, we first verify the reliability of the resolvent-fitting approximation in evaluating the energy and gradient. Then, we apply the theory to compute acrolein's vibrational frequencies and optimize the conical intersections of the retinal model chromophore RPSB6. We also demonstrate the applicability of the current algorithm for computing analytical gradients with many electronic states. Finally, we test the efficiency of the energy and nuclear gradient algorithms with OpenMP/MPI hybrid parallelization.

## 2. THEORY

This section first briefly overviews (X)MCQDPT2[9,37,42] with resolvent fitting and introduces its gradient theory. For readability, we closely follow the notation in the original MCQDPT2 paper by Nakano.[9] Namely, $P$, $Q$, …, $\alpha$, $\beta$, …, $A$, $B$, …, $i$, $j$, …, $a$, $b$, …, $e$, $f$, …, $p$, $q$, … denote the XMCQDPT2 states, reference states, Slater determinants, doubly occupied (inactive) orbitals, active orbitals, virtual orbitals, and general orbitals, respectively. Our implementation is based on Slater determinants but is also applicable for (X)MCQDPT2 implementations based on configuration state functions (CSFs).

**Brief Overview of (X)MCQDPT2.** In (X)MCQDPT2, the electronic Hamiltonian is partitioned into

$$\hat{H} = \hat{H}^{(0)} + \hat{V}. \tag{1}$$

The zeroth-order Hamiltonian is defined as

$$\hat{H}^{(0)} = \hat{P}\hat{f}\hat{P} + \hat{Q}\hat{f}\hat{Q}, \tag{2}$$

where $\hat{P}$ is a projector to the reference space and $\hat{Q}$ is a projector to the expansion basis. The one-electronic Fock operator is defined as

$$\hat{f} = \sum_{pq} f_{pq} \hat{E}_{pq} = \sum_{p} \varepsilon_p \hat{E}_{pp}, \qquad (3)$$

where the orbitals are semicanonicalized (in which the inactive-inactive, active-active and virtual-virtual blocks are diagonalized). Note that although the semicanonical orbitals are the most natural choice for the active orbitals, the first implementation of MCQDPT2 did not stick to the semicanonical active orbitals but also used natural orbitals.[9] The effective Hamiltonian is constructed as

$$H^{\text{eff}}_{\alpha\beta} = \sum_{I} \langle \alpha | \hat{V} | I \rangle \frac{1}{E^{(0)}_{\beta} - E^{(0)}_{I}} \langle I | \hat{V} | \beta \rangle, \qquad (4)$$

where $|I\rangle$ is the determinant out of the CASCI space, and the zeroth-order energies in the denominator are

$$E^{(0)}_{\beta} = \langle \beta | \hat{H}^{(0)} | \beta \rangle, \qquad (5)$$

$$E^{(0)}_{I} = \langle I | \hat{H}^{(0)} | I \rangle. \qquad (6)$$

The effective Hamiltonian is then diagonalized to result in the MCQDPT2 state. In Ref. 9, Nakano presented the explicit formulation of the effective Hamiltonian elements, which reads

$$H_{\alpha\beta}^{\text{eff}} = \sum_{AB} c_{B\alpha} c_{B\beta} \left[ -\left( \sum_{ia'} \frac{2u_{ia'} u_{a'i}}{\varepsilon_{a'} - \varepsilon_i + \Delta E_{B\beta}} \right) + \sum_{ija'b'} \frac{(ia'|jb')[2(a'i|b'j) - (a'j|b'i)]}{\varepsilon_{a'} - \varepsilon_i + \varepsilon_{b'} - \varepsilon_j + \Delta E_{B\beta}} \right]$$

$$+ \sum_{pqB} \langle \alpha | \hat{E}_{pq} | B \rangle c_{B\beta} \left[ \sum_i \frac{u_{iq} u_{pi}}{\varepsilon_p - \varepsilon_i + \Delta E_{B\beta}} - \sum_e \frac{u_{pe} u_{eq}}{\varepsilon_e - \varepsilon_p + \Delta E_{B\beta}} - \sum_{ia'} \frac{u_{ia'}[2(a'i|pq) - (a'q|pi)]}{\varepsilon_{a'} - \varepsilon_i + \varepsilon_p - \varepsilon_q + \Delta E_{B\beta}} \right.$$

$$\left. - \sum_{ia'} \frac{[2(ia'|pq) - (iq|pa')]u_{a'i}}{\varepsilon_{a'} - \varepsilon_i + \Delta E_{B\beta}} + \sum_{ija'} \frac{(ja'|iq)[2(a'j|pi) - (a'i|pj)]}{\varepsilon_{a'} - \varepsilon_j + \varepsilon_p - \varepsilon_i + \Delta E_{B\beta}} - \sum_{ia'b'} \frac{(ia'|pb')[2(a'i|b'q) - (a'q|b'i)]}{\varepsilon_{a'} - \varepsilon_i + \varepsilon_{b'} - \varepsilon_q + \Delta E_{B\beta}} \right]$$

$$+ \sum_{pq,rs,B} \langle \alpha | \hat{E}_{pq,rs} | B \rangle c_{B\beta} \left[ \sum_i \frac{u_{iq}(pi|rs)}{\varepsilon_p - \varepsilon_i + \varepsilon_r - \varepsilon_s + \Delta E_{B\beta}} - \sum_e \frac{u_{pe}(eq|rs)}{\varepsilon_e - \varepsilon_q + \varepsilon_r - \varepsilon_s + \Delta E_{B\beta}} + \sum_i \frac{(iq|rs)u_{pi}}{\varepsilon_p - \varepsilon_i + \Delta E_{B\beta}} \right.$$

$$\left. - \sum_e \frac{(pe|rs)u_{eq}}{\varepsilon_e - \varepsilon_q + \Delta E_{B\beta}} - \frac{1}{2}\sum_{ij} \frac{(iq|js)(pi|rj)}{\varepsilon_p - \varepsilon_i + \varepsilon_r - \varepsilon_j + \Delta E_{B\beta}} - \frac{1}{2}\sum_{a'e} \frac{(pa'|re)(a'q|es)}{\varepsilon_{a'} - \varepsilon_q + \varepsilon_r - \varepsilon_s + \Delta E_{B\beta}} - \frac{1}{2}\sum_{ae} \frac{(pe|ra)(eq|as)}{\varepsilon_e - \varepsilon_q + \varepsilon_a - \varepsilon_s + \Delta E_{B\beta}} \right.$$

$$\left. + \sum_{ia'} \frac{(pa'|iq)(a'i|rs)}{\varepsilon_{a'} - \varepsilon_i + \varepsilon_r - \varepsilon_s + \Delta E_{B\beta}} + \sum_{ia'} \frac{(pa'|is)(a'q|ri)}{\varepsilon_{a'} - \varepsilon_q + \varepsilon_r - \varepsilon_i + \Delta E_{B\beta}} - \sum_{ia'} \frac{(ia'|pq)[2(a'i|rs) - (a's|ri)]}{\varepsilon_{a'} - \varepsilon_i + \varepsilon_r - \varepsilon_s + \Delta E_{B\beta}} \right]$$

$$+ \sum_{pq,rs,tu,B} \langle \alpha | \hat{E}_{pq,rs,tu} | B \rangle c_{B\beta} \left[ \sum_i \frac{(iq|rs)(pi|tu)}{\varepsilon_p - \varepsilon_i + \varepsilon_t - \varepsilon_u + \Delta E_{B\beta}} - \sum_e \frac{(pe|rs)(eq|tu)}{\varepsilon_e - \varepsilon_q + \varepsilon_t - \varepsilon_u + \Delta E_{B\beta}} \right] \quad (7)$$

In this equation, the summation over $a'$ and $b'$ runs over both active and virtual orbitals, $c_{A\alpha}$ is the CI coefficient of the determinant $A$ for the reference state $\alpha$, and **u** is the one-particle perturbation operator defined as

$$u_{pq} = h_{pq} - \varepsilon_p \delta_{pq} + \sum_i 2(pq|ii) - (pi|iq). \quad (8)$$

In the original MCQDPT2, the off-diagonal elements of the zeroth-order Hamiltonian in the state basis are nonzero. This leads to the non-invariance with respect to the rotations among the states, and spurious behavior of the potential energy surfaces near the degeneracy of the reference function. Using the intermediate reference states that diagonalize the zeroth-order Hamiltonian,

$$\langle \alpha | \hat{H}^{(0)} | \beta \rangle = \delta_{\alpha\beta} E_\beta^{(0)}, \quad (9)$$

can address this problem. This modified version of MCQDPT2 is the extended MCQDPT2 (XMCQDPT2),[37] which was also applied to MS-CASPT2, resulting in the XMS-CASPT2 method.[15] An intermediate theory between MS-CASPT2 and XMS-CASPT2, i.e., extended dynamically weighted CASPT2, was also suggested, and it has been demonstrated that the theory performs well for geometries with large and small separations between electronic states.[43]

Multiconfiguration perturbation theories usually suffer from the intruder state problem.[44-46] The intruder state is present when the denominator $\Delta_I$ in the (X)MCQDPT2 expansion is close to zero such that

$$H_{\alpha\beta} = \sum_I \frac{N_I}{\Delta_I} \tag{10}$$

diverges. This problem can be circumvented by the intruder state avoidance (ISA) technique.[44] With the ISA technique, the denominator is regularized as

$$\Delta_I \to \Delta_I + \frac{\tau}{\Delta_I}, \tag{11}$$

where $\tau$ is the ISA parameter with a dimension $E_h^2$. We note that this corresponds to the imaginary shift technique in CASPT2.[20,46]

**Resolvent Fitting.** We now review the resolvent-fitting (table-driven) approximation in the (X)MCQDPT2 calculations.[38] This method considers the perturbation summations over the Slater determinants (resolvents) as a smooth function of the zeroth-order energy difference. This is always true when the ISA technique is employed. Then, the terms required for perturbation summations are interpolated rather than explicitly computed. We take the first term in the (X)MCQDPT2 expansion in Eq. (7) as an example:

$$H_{\alpha\beta}^{\text{eff}} = \sum_B c_{B\alpha} c_{B\beta} \left[ -\sum_{ia'} \frac{2\Delta_{a'i,B\beta} u_{ia'} u_{a'i}}{\Delta_{a'i,B\beta}^2 + \tau} + \sum_{ia'jb'} \frac{\Delta_{a'ib'j,B\beta}(ia' | jb')[2(a'i|b'j)-(a'j|b'i)]}{\Delta_{a'ib'j,B\beta}^2 + \tau} \right], \tag{12}$$

where the denominators are

$$\Delta_{a'i,B\beta} = \varepsilon_{a'} - \varepsilon_i + \Delta E_{B\beta}, \tag{13}$$

$$\Delta_{a'ib'j,B\beta} = \varepsilon_{a'} - \varepsilon_i + \varepsilon_{b'} - \varepsilon_j + \Delta E_{B\beta}. \tag{14}$$

In the original (X)MCQDPT2, the latter summation is explicitly calculated. For brevity, we rewrite the latter sum as

$$S^{(0)}(\Delta E_{B\beta}) = -\sum_{ia'} \frac{2\Delta_{a'i,B\beta} u_{ia'} u_{a'i}}{\Delta^2_{a'i,B\beta} + \tau} + \sum_{ia'jb'} \frac{\Delta_{a'ib'j,B\beta}(ia'|jb')[2(a'i|b'j) - (a'j|b'i)]}{\Delta^2_{a'ib'j,B\beta} + \tau}. \tag{15}$$

Then, one can rewrite the contribution of this term to the effective Hamiltonian as

$$H^{\text{eff}}_{\alpha\beta} = \sum_B c_{B\alpha} c_{B\beta} S^{(0)}(\Delta E_{B\beta}). \tag{16}$$

Granovsky realized that this $S^{(0)}(\Delta E_{B\beta})$ can be regarded as a smooth function of $\Delta E_{B\beta}$ when the ISA technique is applied. Thus, it is possible to interpolate this function.[38] For interpolation, a helper grid $\lambda$ is introduced. At each grid point $\lambda_g$, $S^{(0)}(\lambda_g)$ is evaluated as

$$S^{(0)}(\lambda_g) = -\sum_{ia'} \frac{2\Delta_{a'i,g} u_{ia'} u_{a'i}}{\Delta^2_{a'i,g} + \tau} + \sum_{ia'jb'} \frac{\Delta_{a'ib'j,g}(ia'|jb')[2(a'i|b'j) - (a'j|b'i)]}{\Delta^2_{a'ib'j,g} + \tau}, \tag{17}$$

where $\Delta_{a'i,g} = \varepsilon_{a'} - \varepsilon_i + \lambda_g$ and $\Delta_{a'ib'j,g} = \varepsilon_{a'} - \varepsilon_i + \varepsilon_{b'} - \varepsilon_j + \lambda_g$. The value of $S^{(0)}$ is saved for each $\lambda_g$. The separation of $\Delta\lambda = 0.05 E_h$ is sufficient to achieve accuracy on the order of $1.0 \times 10^{-7} E_h$. In our implementation, the values of $\lambda$ at all geometrical points are separated by 0.05 $E_h$ starting from zero (e.g., the values of $\lambda$ are …,–0.05, 0.00, +0.05 …). The maximum ($\lambda_{\max}$) and minimum ($\lambda_{\min}$) values of $\lambda$ are chosen to cover all possible values of $\Delta E_{B\beta}$ with eight-point polynomial interpolation. The total number of $\lambda$ is $N_{\text{grid}}$, which is simply $N_{\text{grid}} = \frac{\lambda_{\max} - \lambda_{\min}}{0.05} + 1$. When evaluating the effective Hamiltonian, we loop over $\alpha$, $\beta$, and $B$. The interpolant $S^{(0)}(\Delta E_{B\beta})$ can be approximated by the Lagrangian interpolation as

$$S^{(0)}(\Delta E_{B\beta}) \approx \sum_g S^{(0)}(\lambda_g) W_g(\Delta E_{B\beta}), \tag{18}$$

where the interpolation weights $W_g$ can be determined by arbitrary means. In our implementation, we use the eight-point polynomial interpolation, which means that $N_{\text{fgrid}} = 8$ for each $\Delta E_{B\beta}$. The weights are

$$W_g(\Delta E_{B\beta}) = \prod_{h \neq g} \frac{\Delta E_{B\beta} - \lambda_h}{\lambda_g - \lambda_h}. \tag{19}$$

For computational efficiency, the interpolation coefficients are evaluated and stored before the (X)MCQDPT2 calculation. The contributions of the zero-particle term to the effective Hamiltonian is then

$$\begin{aligned}
H_{\alpha\beta}^{\text{eff},0} &= \sum_B c_{B\alpha} c_{B\beta} \sum_g W_g(\Delta E_{B\beta}) S^{(0)}(\lambda_g) \\
&= \sum_g \left[ \sum_B c_{B\alpha} c_{B\beta} W_g(\Delta E_{B\beta}) \right] S^{(0)}(\lambda_g) \\
&\equiv \sum_g P_g^{\alpha\beta} S^{(0)}(\lambda_g),
\end{aligned} \tag{20}$$

where we have defined $P_g^{\alpha\beta}$, which is useful in formulating the analytical gradient theory in the following. Now, this can be applied to one- to three- particle terms by rewriting Eq. (7) as

$$\begin{aligned}
H_{\alpha\beta}^{\text{eff}} &= \sum_{AB} c_{A\alpha} c_{B\beta} S^{(0)}(\Delta E_{B\beta}) + \sum_{pqB} \langle \alpha | \hat{E}_{pq} | B \rangle c_{B\beta} S_{pq}^{(1)}(\Delta E_{B\beta}) \\
&+ \sum_{pq,rs,B} \langle \alpha | \hat{E}_{pq,rs} | B \rangle c_{B\beta} S_{pq,rs}^{(2)}(\Delta E_{B\beta}) \\
&+ \sum_{pq,rs,tu,B} \langle \alpha | \hat{E}_{pq,rs,tu} | B \rangle c_{B\beta} S_{pq,rs,tu}^{(3)}(\Delta E_{B\beta}) \\
&\equiv H_{\alpha\beta}^{\text{eff},0} + H_{\alpha\beta}^{\text{eff},1} + H_{\alpha\beta}^{\text{eff},2} + H_{\alpha\beta}^{\text{eff},3}.
\end{aligned} \tag{21}$$

For example, the one-particle term ($H_{\alpha\beta}^{\text{eff},1}$) can be similarly rewritten as

$$H_{\alpha\beta}^{\text{eff},1} = \sum_B \sum_{ab} \langle \alpha | \hat{E}_{ab} | B \rangle c_{B\beta} \sum_g W_g(\Delta E_{B\beta}) S_{ab}^{(1)}(\lambda_g)$$

$$= \sum_{ab} \sum_g \left[ \sum_B \langle \alpha | \hat{E}_{ab} | B \rangle c_{B\beta} W_g(\Delta E_{B\beta}) \right] S_{ab}^{(1)}(\lambda_g). \quad (22)$$

$$\equiv \sum_{ab} \sum_g P_{ab,g}^{\alpha\beta} S^{(1)}(\lambda_g).$$

Then, its contribution to the energy of the (X)MCQDPT2 state $P$ is

$$E_P = \sum_{ab} \sum_g \sum_{\alpha\beta} T_{\alpha\beta}^P P_{ab,g}^{\alpha\beta} S^{(1)}(\lambda_g) \quad (23)$$

where $T_{\alpha\beta}^P$ is defined as $T_{\alpha\beta}^P = R_{\alpha P} R_{\beta P}$ and **R** is the eigenvector of the effective Hamiltonian. We further define

$$P'_{ab,g} = \sum_{\alpha\beta} T_{\alpha\beta}^P P_{ab,g}^{\alpha\beta}$$

$$= \sum_{\alpha\beta} T_{\alpha\beta}^P \sum_B \langle \alpha | \hat{E}_{ab} | B \rangle c_{B\beta} W_g(\Delta E_{B\beta}). \quad (24)$$

The evaluation of **P'** requires $N_{\text{act}}^{2M} N_{\text{det}} N_{\text{grid}} N_{\text{state}}^2$ operations for the *M*-particle term. This quantity is directly related to the (X)MCQDPT2 energy and is valuable in analytical gradient theory.

Now, let us discuss the savings associated with the resolvent-fitting approximation. Without resolvent-fitting approximation, an evaluation of the second term in Eq. (12) requires $N_{\text{vir}}^2 N_{\text{core}}^2 N_{\text{det}} N_{\text{state}}^2$. One evaluates this term with resolvent-fitting by first evaluating the $S^{(0)}$ values at the grid points [Eq. (17), $N_{\text{grid}} N_{\text{vir}}^2 N_{\text{core}}^2$], interpolating it at all $\Delta E_{B\beta}$ values [Eq. (18), $N_{\text{fgrid}} N_{\text{state}} N_{\text{det}}$], and then contracting it with the CI coefficients [Eq. (16), $N_{\text{state}}^2 N_{\text{det}}$]. The combined count of operations $N_{\text{grid}} N_{\text{vir}}^2 N_{\text{core}}^2 + (N_{\text{fgrid}} + N_{\text{state}}) N_{\text{state}} N_{\text{det}}$ is always smaller than $N_{\text{vir}}^2 N_{\text{core}}^2 N_{\text{det}} N_{\text{state}}^2$ for the calculations without the resolvent-fitting approximation, except for the case that $N_{\text{det}}$ and $N_{\text{state}}$ are very small. Therefore, the resolvent-fitted (X)MCQDPT2 is almost always favorable over the canonical (X)MCQDPT2, particularly when the numbers of determinants and states are large (see also numerical examples). Similarly, the most expensive 3-particle term [$H_{\alpha\beta}^{\text{eff},3}$ in Eq. (21)] requires $N_{\text{grid}}(N_{\text{core}}+N_{\text{vir}})N_{\text{act}}^6 + (N_{\text{fgrid}}+N_{\text{state}})N_{\text{act}}^6 N_{\text{det}} N_{\text{state}}$ and $N_{\text{grid}}(N_{\text{core}}+N_{\text{vir}})N_{\text{act}}^6 N_{\text{det}} N_{\text{state}}^2$ operations with and without the resolvent-fitting approximation (the XMS-CASPT2 and QD-NEVPT2 analytical gradient algorithms require $\frac{1}{8}N_{\text{state}}^2 N_{\text{det}} N_{\text{act}}^8$ operations in 4RDM evaluations). Therefore, in principle, the cost for the (X)MCQDPT2 gradient can be comparable to that of the XMS-CASPT2 or QD-NEVPT2 gradients.

**Analytical Gradient Theory.** Let us now derive the expressions for the pseudodensities and CI derivatives for the resolvent-fitted (X)MCQDPT2, which are necessary for evaluating the analytical gradient. The zero-particle energy is

$$E_P = \sum_{\alpha\beta} T_{\alpha\beta}^P \sum_B c_{B\alpha} c_{B\beta} \sum_g S^{(0)}(\lambda_g) W_g$$
$$= \sum_g P'_g S^{(0)}(\lambda_g). \tag{25}$$

Here, we have defined

$$P'_g = \sum_{\alpha\beta} T_{\alpha\beta}^P \sum_B c_{B\alpha} c_{B\beta} W_g. \tag{26}$$

The expressions for the CI derivative are just

$$y_{B\alpha} = \sum_\beta T_{\alpha\beta}^P c_{B\beta} \sum_g S^{(0)}(\lambda_g) W_g, \tag{27}$$

$$y_{B\beta} = \sum_\alpha T_{\alpha\beta}^P c_{B\alpha} \sum_g S^{(0)}(\lambda_g) W_g. \tag{28}$$

The density matrices are also simply obtained. If one uses fixed grid points, the derivative with respect to $\lambda$ is zero. Consequently, the weighting factor is solely a function of $\Delta E_{B\beta}$. This means that the density matrix elements, except those from $\Delta E_{B\beta}$, have contributions only from $S^{(0)}$. For denominators, we write the Lagrangian as

$$\mathcal{L}_{\text{XMCQDPT2}} = E_P + Q^{\tau}_{a'ig}(\varepsilon_{a'} - \varepsilon_i + \lambda_g - \Delta_{a'i,g}) + Q^{\tau}_{a'ib'jg}(\varepsilon_{a'} - \varepsilon_i + \varepsilon_{b'} - \varepsilon_j + \lambda_g - \Delta_{a'ib'j,g}). \qquad (29)$$

Then, the multipliers $Q^{\tau}_{a'ig}$ and $Q^{\tau}_{a'ib'jg}$ are

$$Q^{\tau}_{a'ig} = \frac{\Delta^2_{a'i,g} - \tau}{\Delta^2_{a'i,g} + \tau} P'_g, \qquad (30)$$

$$Q^{\tau}_{a'ib'jg} = \frac{\Delta^2_{a'ib'j,g} - \tau}{\Delta^2_{a'ib'j,g} + \tau} P'_g. \qquad (31)$$

If $\lambda$ is fixed at all molecular geometries (which is the case in our implementation), the contributions of the zero-particle term to the density matrix elements are

$$\begin{aligned}
d^{\text{Fock}}_{a'a'} &= \sum_g Q^{\tau}_{a'ig} + \sum_g Q^{\tau}_{a'ib'jg}, \\
d^{\text{Fock}}_{b'b'} &= \sum_g Q^{\tau}_{a'ib'jg}, \\
d^{\text{Fock}}_{ii} &= -\sum_g Q^{\tau}_{a'ig} - \sum_g Q^{\tau}_{a'ib'jg}. \\
d^{\text{Fock}}_{jj} &= -\sum_g Q^{\tau}_{a'ib'jg}.
\end{aligned} \qquad (32)$$

The density matrix elements from the numerator are

$$\begin{aligned}
d^u_{ia'} &= -\sum_g P'_g \frac{u_{a'i}}{\varepsilon_{a'} - \varepsilon_i + \lambda_g} \\
d^u_{a'i} &= -\sum_g P'_g \frac{u_{ia'}}{\varepsilon_{a'} - \varepsilon_i + \lambda_g} \\
D^{(1)}_{ia'jb'} &= \sum_g P'_g \frac{2(a'i | b'j) - (a'j | b'i)}{\varepsilon_{a'} - \varepsilon_i + \varepsilon_{b'} - \varepsilon_j + \lambda_g}, \\
D^{(1)}_{a'ib'j} &= \sum_g P'_g \frac{2(ia' | jb')}{\varepsilon_{a'} - \varepsilon_i + \varepsilon_{b'} - \varepsilon_j + \lambda_g}, \\
D^{(1)}_{a'jb'i} &= -\sum_g P'_g \frac{(ia' | jb')}{\varepsilon_{a'} - \varepsilon_i + \varepsilon_{b'} - \varepsilon_j + \lambda_g}.
\end{aligned} \qquad (33)$$

As we interpolate the *S*-values, the derivatives of the weights should contribute as

$$G_{B\beta} = \sum_{\alpha\beta} T^P_{\alpha\beta} \sum_B c_{B\alpha} c_{B\beta} \sum_g S^{(0)}(\lambda_g) \frac{\partial W_g}{\partial \Delta E_{B\beta}}, \qquad (34)$$

and they can be formed simultaneously with the CI derivatives. This quantity then modifies the density matrices as

$$d^{\text{Fock}}_{aa} = G_{B\beta} \eta^B_a, \qquad (35)$$

$$d^{\text{Fock}}_{ab} = -G_{B\beta} \langle \beta | \hat{E}_{ab} | \beta \rangle, \qquad (36)$$

where $\eta^B_a$ is the occupation number of orbital $a$ in determinant $B$. The analytical expression for $\partial W_g / \partial \Delta E_{B\beta}$ is

$$\frac{\partial W_g}{\partial \Delta E_{B\beta}} = \left( \prod_{h \neq g} \frac{1}{\lambda_g - \lambda_h} \right) \times \left( \sum_{l \neq g} \frac{\prod_{h \neq g} \Delta E_{B\beta} - \lambda_h}{\Delta E_{B\beta} - \lambda_l} \right), \qquad (37)$$

but in our working implementation, we have employed the numerical differentiation of $W_g$ with respect to $\Delta E_{B\beta}$, which is way more efficient than the analytical derivative without a loss of accuracy (the error in the final gradient is below $1.0 \times 10^{-10}$ $E_h$ bohr$^{-1}$). We store this derivative before the (X)MCQDPT2 gradient calculations, where the storage requirement is just the same as one for $W_g(\Delta E_{B\beta})$. With all these terms, one can evaluate the derivatives of the Lagrangian with respect to the orbital parameters as

$$\mathbf{Y} = 2 \left[ \mathbf{h}\mathbf{d}^{(0)} + \mathbf{u}\mathbf{d}^u + \mathbf{f}\mathbf{d}^{\text{Fock}} + \mathbf{g}(\mathbf{d}^u)\mathbf{d}^{\text{clo}} + \mathbf{g}(\mathbf{d}^{\text{Fock}})\mathbf{d}^{(0)} + \sum_{kl} \mathbf{D}^{kl} \mathbf{K}^{lk} \right], \qquad (38)$$

where $\mathbf{d}^{(0)}$ and $\mathbf{D}^{(0)}$ are the one- and two-electron density matrices of the reference function, $D^{kl}_{rs} = \frac{1}{2}\left[ D^{(0)}_{rkls} + D^{(1)}_{rkls} \right]$, $\mathbf{d}^{\text{clo}}$ is the one-electron density matrix only for the closed orbitals, and $K_{rs}^{lk} = (rl \mid sk)$. The derivatives **Y** and **y** are then used in solving the so-called Z-vector equation in the same manner as for the canonical XMCQDPT2 analytical gradient.[30]

**Density Fitting in Y Evaluation.** An evaluation of the two-particle density matrix, **D**, is necessary for gradient evaluation. Unfortunately, this matrix is $(N_{act} + N_{core})^2 (N_{vir} + N_{act})^2$ in size, being 80 GB when $N_{act} + N_{core} = 100$ and $N_{vir} + N_{act} = 1000$. This makes the algorithm (unnecessarily) costly in terms of memory use. We have avoided the storage of the full **D** by exploiting the density-fitting algorithm in evaluating the last term in Eq. 38, i.e., $\sum_{kl} \mathbf{D}^{kl} \mathbf{K}^{lk}$.

One can evaluate the $\sum_{kl} \mathbf{D}^{kl} \mathbf{K}^{lk}$ term as in Eqs. 24 to 30 in Ref. 16. The **D** tensor is contracted with the density-fitting integral (Eq. 28 in Ref. 16) to form the intermediate tensor **V** as

$$V_{si}^P = \sum_{jt} D_{isjt} \sum_Q J_{PQ}^{-1}(Q \mid jt) \equiv \sum_{jt} D_{isjt} d_{jt}^P, \quad (39)$$

where **J** is the electron repulsion integral over an auxiliary basis, a central quantity for using the Coulomb metric in density fitting.[47] The zero-particle terms (Eq. 33) contribute to the largest subtensor to **D**. The full zero-particle part of **D** has $N_{core}^2 (N_{vir} + N_{act})^2$ elements. Instead of storing the entire matrix, we loop over $i$ and $j$ and allocate the submatrix with fixed $\{i, j\}$, the size of which is only $(N_{act} + N_{vir})^2$. Similarly, for the last term in the third line in Eq. 7, we loop over $a'$ and $b'$, and allocate the submatrix with fixed $\{a', b'\}$, whose size is $N_{occ}^2$. Then, one forms tensor **V** by performing the contraction in Eq. 39 separately for each pair $\{i, j\}$ and $\{a', b'\}$. Then, the size of the largest **D** matrix stored in the memory is $(N_{act} + N_{core})^2 (N_{vir} + N_{act}) N_{act}$ for two- and three-particle terms. Overall, storing only submatrices of **D** results in the saving of memory by a factor of $\frac{N_{act}}{N_{act}+N_{vir}}$. When a large basis set is used, $N_{vir} \gg N_{act}$, which results in large savings in terms of memory.

## 3. NUMERICAL EXAMPLES

This section presents numerical examples of the resolvent-fitted XMCQDPT2 analytical gradient theory. First, we check the fidelity of the resolvent-fitting approximation in evaluating energy and gradients by comparing the results with the canonical (X)MCQDPT2 and the accuracy of analytical gradients. Then, we optimize the equilibrium geometries of acrolein and evaluate the vibrational frequencies at them (with finite-difference differentiation of the first-order derivatives) using the XMCQDPT2 and XMS-CASPT2 theories. We also present the optimization of the conical intersection geometries in RPSB6 with the (12*e*, 12*o*) active space and compare its accuracy against the XMS-CASPT2 results. We then demonstrate how the current algorithm performs for systems with many electronic states and the parallel efficiency in and across computer nodes. The density-fitting approximation is used in all calculations, using the JKFIT basis functions. In all XMCQDPT2 calculations, we used an ISA parameter of 0.02 $E_h^2$.[44]

**Accuracy and Efficiency of Resolvent Fitting.** Let us discuss the accuracy of the resolvent-fitting approximation in evaluating energy, gradient, and optimizing molecular geometries. We have tested four systems: LiF, truncated rhodopsin protonated Schiff base model penta-2,4-dieniminium cation (PSB3), benzene, and the green fluorescent protein chromophore model *para*-hydroxybenzilideneimidazolin-5-one (*p*HBI) anion. See Scheme 1 for the structures of PSB3 and *p*HBI. In Table 1, we compare the absolute value of the gradient for LiF and the norms of the gradients at the CASSCF equilibrium geometries for others. We also compare the resulting energy at the test geometry (LiF) and the optimized geometries (PSB3, benzene, pHBI).

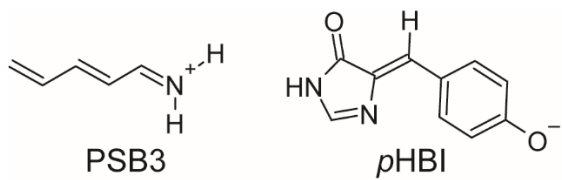

**Scheme 1.** Structures of PSB3 and pHBI anion.

**Table 1.** Comparisons of the XMCQDPT2 energy (in $E_h$) and gradient (in $E_h$ bohr$^{-1}$) with and without resolvent-fitting approximation for LiF, PSB3, benzene, and pHBI anion.

| System[a] | Active and model space | State | No resolvent fitting | With resolvent fitting |
|---|---|---|---|---|
| LiF[b] | (6e,4o) 4×4 | $S_0$ | -106.917607709**8** | -106.917607708**9** |
| | | $S_1$ | -106.8589106156 | -106.8589106156 |
| | | $S_2$ | -106.8589106156 | -106.8589106156 |
| | | $S_3$ | -106.846177538**7** | -106.846177539**7** |
| | | gradient[c] | 0.034438841**8** | 0.034438841**6** |
| PSB3[e] | (6e,6o) 3×3 | $S_0$ | -249.053576707**8** | -249.053576708**0** |
| | | $S_1$ | -248.90480279**11** | -248.90480279**40** |
| | | $S_2$ | -248.85922715**07** | -248.85922714**75** |
| | | gradient[d] | 0.042225349**8** | 0.042225350**3** |
| benzene[e] | (6e,6o) 1×1 | $S_0$ | -231.513447904**0** | -231.513447904**3** |
| | | gradient[d] | 0.023224755**8** | 0.023224765**0** |
| pHBI anion[e] | (4e,3o) 3×3 | $S_0$ | -643.4545754**394** | -643.4545754**572** |
| | | $S_1$ | -643.3624840**303** | -643.3624840**064** |
| | | $S_2$ | -643.3124**084484** | -643.3124**035667** |
| | | gradient[d] | 0.1240335**551** | 0.1240335**761** |

[a]the def2-SVP/def2-SVP-JKFIT basis set and the cc-pVDZ/cc-pVDZ-JKFIT basis set were used for LiF and the rest, respectively. [b]values at the test geometry, where two nuclei were separated by 6.0 bohr. [c]The $S_0$ gradient value along the Z-axis. [d]norm of the $S_0$ gradients at the SA-CASSCF $S_0$ optimized geometry. [e]energies at the XMCQDPT2 $S_0$ optimized geometry.

The errors of the energies are all below $10^{-5}$ $E_h$. The $S_2$ energy of $p$HBI anion shows the largest error of $4.9\times10^{-6}$ $E_h$. Note that the determinant (CI) space is only 9 for this system, and the computational cost reduction by the resolvent-fitting approximation is relatively small (see below). When the determinant space is larger, the error is below $1.0\times10^{-8}$ $E_h$ for all cases. The computed analytical gradients also had errors of similar magnitudes. In terms of norms, the errors were within $1.0\times10^{-8}$ a.u. ($E_h$ bohr$^{-1}$) for LiF, PSB3, and benzene, and within $1.0\times10^{-7}$ a.u. for $p$HBI anion. We have shown in our previous work[30] that the analytical and numerical gradients agree with the error below $5.0\times10^{-6}$ a.u. in the canonical XMCQDPT2 theory. This is also the case for the resolvent-fitted XMCQDPT2. All the optimized geometries with and without resolvent fitting approximation were within $10^{-5}$ Å (in terms of root-mean-squared Euclidean distances), which is natural from small errors in analytical gradients. These results agree well with the previous validations of resolvent-fitting approximations in terms of energy.[38] Overall, we can conclude that the resolvent-fitted XMCQDPT2 yields very similar results to the canonical XMCQDPT2 in terms of both energies and gradients. At the same time, some caution is needed when the determinant (CI) space is small.

Next, let us compare the computational cost with and without resolvent-fitting approximations. In Table 2, we present the wall times with and without the resolvent-fitting approximations. In all the cases, there is a significant improvement over the canonical XMCQDPT2 theory. In particular, the wall time for evaluating the resolvent-fitted gradient of PSB3 is about 30 times shorter than one for the canonical gradient. The computational cost is more saved when the number of states is larger. While the active space size is the same as PSB3, the resolvent-fitting approximation reduced the wall time for evaluating gradients of single-state benzene by a factor of only ~5. When the dimension of the CI space is small (the $p$HBI anion case), the gain in the efficiency is also relatively small. Both results are natural, as the term dependent on the number of states is decoupled from the term dependent on the number of orbitals or determinants by the resolvent-fitting approximation.

**Table 2.** Comparison of wall time (in seconds) with and without resolvent-fitting approximation for PSB3, benzene, and $p$HBI anion.[a]

| System[b] | Active and model space | | energy[c] | gradient[d] | Z-vector, contraction[e] | total |
|---|---|---|---|---|---|---|
| PSB3 | (6$e$,6$o$) 3×3 | No resolvent fitting | 26 | 285 | 4 | 315 |
| | | Resolvent fitting | 1 | 7 | 4 | 12 |
| benzene | (6$e$,6$o$) 1×1 | No resolvent fitting | 3 | 40 | 4 | 47 |
| | | Resolvent fitting | 1 | 5 | 4 | 10 |
| $p$HBI anion | (4$e$,3$o$) 3×3 | No resolvent fitting | 3 | 208 | 32 | 243 |
| | | Resolvent fitting | 2 | 23 | 32 | 57 |

[a] These times are measured using 18 physical cores in an Intel Xeon Gold 6140 processor (2.3 GHz). [b]cc-pVDZ/cc-pVDZ-JKFIT basis set was used for all the calculations. [c]Wall time for evaluating MCQDPT2 energy. [d]Wall time for evaluating pseudodensities and CI derivatives. [e]Wall time for solving the Z-vector equation and contraction with the integral derivatives.

Overall, there are no significant errors introduced by the resolvent-fitting approximation (at least for the test cases we have selected) in terms of energy, gradient, and optimized geometries. When the sizes of active space or model space are large, the resolvent-fitting approximation reduces the computational cost almost by order of magnitude. These results reconfirm the reliability of the resolvent-fitting approximation, along with the previous benchmark by Granovsky.[38] Therefore, we highly recommended using the resolvent-fitting approximation when the ISA parameter is nonzero.

**Geometry Optimizations and Frequencies of Acrolein.** Employing the internally contracted basis functions offers an advantage in terms of computational cost. However, there can be some artifacts of using them, apart from missing contraction energies[31-34] or treating states with small energetical gaps. Here, we show a simple example of an artifact in geometry optimizations and obtaining vibrational frequencies using finite-difference differentiation.

**Table 3.** Ten lowest eigenvalues of the mass-weighted Hessian (vibrational frequency) in cm$^{-1}$. The translational and rotational components are projected out from the Hessian.

| | | Overlap Threshold | Equilibrium $E$ | 1 | 2 | 3 | 4 | 5 | 6 | 7 | 8 | 9 | 10 |
|---|---|---|---|---|---|---|---|---|---|---|---|---|---|
| $S_0$ | CASSCF | | -190.8247072 | 0 | 0 | 0 | 0 | 0 | 0 | 188 | 304 | 574 | 722 |
| | XMS-CASPT2(D) SS-SR | $1.0\times10^{-9}$ | -191.3654421 | **-68** | 0 | 0 | 0 | 0 | 0 | 167 | 294 | 679 | 769 |
| | | $1.0\times10^{-8}$ | -191.3654412 | **-65** | 0 | 0 | 0 | 0 | 0 | 187 | 294 | 539 | 679 |
| | | $1.0\times10^{-7}$ | -191.3654398 | 0 | 0 | 0 | 0 | 0 | 0 | 183 | 295 | 613 | 679 |
| | XMS-CASPT2 SS-SR | $1.0\times10^{-9}$ | -191.3628701 | **-14** | 0 | 0 | 0 | 0 | 0 | 199 | 293 | 660 | 679 |
| | | $1.0\times10^{-8}$ | -191.3628692 | **-378** | 0 | 0 | 0 | 0 | 0 | 101 | 292 | 449 | 680 |
| | | $1.0\times10^{-7}$ | -191.3628679 | 0 | 0 | 0 | 0 | 0 | 0 | 181 | 293 | 607 | 679 |
| | XMS-CASPT2 MS-MR | $1.0\times10^{-9}$ | -191.3631988 | **-12** | 0 | 0 | 0 | 0 | 0 | 122 | 293 | 387 | 679 |
| | | $1.0\times10^{-8}$ | | (convergence not reached) | | | | | | | | | |
| | | $1.0\times10^{-7}$ | -191.3631959 | 0 | 0 | 0 | 0 | 0 | 0 | 175 | 293 | 561 | 679 |
| | XMCQDPT2 | | -191.3668636 | 0 | 0 | 0 | 0 | 0 | 0 | 181 | 294 | 565 | 678 |
| $S_1$ [a] | CASSCF | | -190.7210864 | 0 | 0 | 0 | 0 | 0 | 0 | 260 | 368 | 534 | 558 |
| | XMS-CASPT2(D) SS-SR | $1.0\times10^{-7}$ | -191.2497432 | 0 | 0 | 0 | 0 | 0 | 0 | 235 | 370 | 503 | 604 |
| | XMS-CASPT2 SS-SR | $1.0\times10^{-7}$ | -191.2479949 | 0 | 0 | 0 | 0 | 0 | 0 | 234 | 365 | 504 | 604 |
| | XMS-CASPT2 MS-MR | $1.0\times10^{-7}$ | -191.2482728 | 0 | 0 | 0 | 0 | 0 | 0 | 234 | 398 | 506 | 604 |
| | XMCQDPT2 | | -191.2499739 | 0 | 0 | 0 | 0 | 0 | 0 | 235 | 366 | 504 | 604 |

[a] Convergence problem for CASPT2 calculations with the threshold of $1.0\times10^{-9}$ or $1.0\times10^{-8}$.

We have optimized the $S_0$ and $S_1$ geometries of acrolein and computed finite-difference Hessian at the equilibrium geometries using the CASSCF, XMS-CASPT2 (with both "SS-SR" and "MS-MR" internal contraction schemes), and XMCQDPT2 methods. We used the cc-pVDZ basis set. We employed the imaginary shift of 0.2 $E_h$ for all the XMS-CASPT2 calculations. The active space included all the π orbitals, which resulted in (6$e$,5$o$) space. We employed the model space of 3×3.

When constructing an internally contracted basis, one should get rid of the linear dependencies between them. In many practical implementations, one removes the linear dependence in the internally contracted basis by deleting all the eigenvalues of the overlap matrix whose absolute values are below the threshold value. For example, the default threshold value is $1.0\times10^{-9}$ for BAGEL (keyword: `thresh_overlap`), and $1.0\times10^{-8}$ for MOLCAS (keyword: `THREsholds`) and MOLPRO (keyword: `throvl` in `gthresh`). However, these values are somewhat arbitrary and are not universally applicable for all molecular systems to obtain a consistent internally contracted basis for different molecular geometries.

Due to the inconsistency in the internal basis in distinct conformations, two problems arise for acrolein. First, with the overlap threshold of $1.0\times10^{-9}$ or $1.0\times10^{-8}$ are employed, we encountered a convergence problem when optimizing $S_1$ equilibrium structures. The optimization converged when the overlap threshold is $1.0\times10^{-7}$. Second, there is a spurious negative frequency at the equilibrium geometries even when convergence is reached (see the resulting ten lowest eigenvalues from the finite-difference Hessians in Table 3). In all XMS-CASPT2 results, the lowest frequency is nonzero when the overlap threshold is $1.0\times10^{-9}$ or $1.0\times10^{-8}$. Of course, one can solve this problem by implementing and calculating analytical Hessian for CASPT2 methods, but it would be challenging due to the complexity of the CASPT2 formulation. The XMCQDPT2 calculations are free from such problems and can be valuable alternatives to the CASPT2 calculations when the internally contracted basis poses issues.

**Minimum Energy Conical Intersection Optimizations of RPSB6.** An investigation of photoisomerization of the retinal protonated Schiff base (RPSB) is essential in vision studies.[48-51] Many truncated models of the RPSB have been employed to inspect the photoisomerization of rhodopsin.[48-51] Two photodeactivation pathways are considered responsible: the twisting of the 11–12 and 13–14 double bonds. In the author and Shiozaki's previous benchmarks using XMS-CASPT2 with the cc-pVDZ basis set,[52] the 13–14 minimum energy conical intersection (MECI) in RPSB6 was thermally unreachable at the Franck–Condon point. Here, we present the resolvent-fitted XMCQDPT2 optimization of the MECIs in RPSB6 using the cc-pVTZ basis set[53] and its corresponding JKFIT basis set,[54] which resulted in 1108 basis functions. We include the three lowest singlet states in the SA-CASSCF and XMCQDPT2 calculations. We perform MECI searches with the gradient projection method.[55] We note that we could not afford the computational cost with cc-pVTZ in our previous XMS-CASPT2 benchmark, and the XMS-CASPT2/cc-pVDZ results are compared.

The structures of the resulting all-*trans* $S_0$ equilibrium geometry, i.e., the 11–12 and 13–14 MECIs, are shown in Figure 1. First, the vertical excitation energy is 2.14 eV, which is approximately 0.32 eV lower than the XMS-CASPT2/cc-pVDZ value (2.46 eV). To check the main reason behind this discrepancy, we optimize the $S_0$ geometry with XMCQDPT2/cc-pVDZ. The resulting vertical excitation energy is 2.27 eV, which is 0.13 eV higher than the cc-pVTZ value. This result means that the XMCQDPT2 vertical excitation energy of RPSB6 is very sensitive to the basis set selection. The geometry optimization with XMS-CASPT2(D)/cc-pVTZ yields vertical excitation energies of 2.18 eV and 2.41 eV with an imaginary shift of 0.2 $E_h$ and a real shift of 0.5 $E_h$, respectively. This implies that the vertical excitation energy is highly shift-dependent as well. The previous results on various shifts in CASPT2 calculations show that a high real shift leads to severe changes in the computational results. Therefore, lower values of the imaginary shift (equivalent to the ISA parameter in XMCQDPT2) are recommended.[20,46] Our selection of the ISA parameter corresponds to the imaginary shift of 0.14 $E_h$, and we can assume that the XMCQDPT2 results are closer to the zero-shift limit.

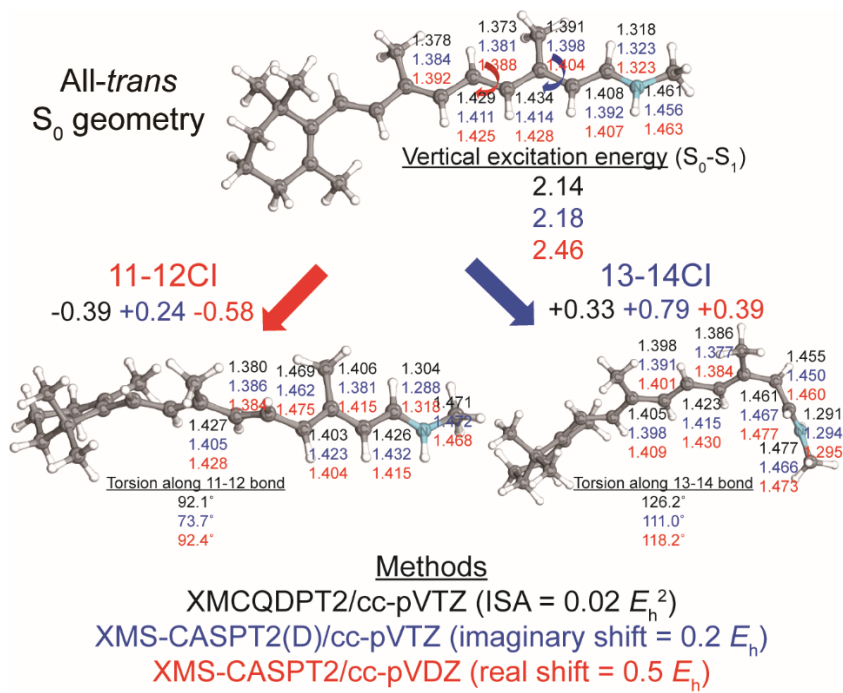

**Figure 1.** Optimized all-*trans* S$_0$ and MECI structures of RPSB6. All bond lengths are shown in Å. For comparative purposes, the XMS-CASPT2(D)/cc-pVTZ geometrical parameters are shown in blue, and XMS-CASPT2/cc-pVDZ geometrical parameters from Ref. 52 are shown in red. The energies with respect to the S$_1$ energy at the Franck–Condon point (in eV) are also shown. The molecular graphics were generated using the IboView software.[56,57]

The energies of the 11–12 MECI and 13–14 MECI are 0.39 eV lower and 0.33 eV higher than the S$_1$ energy at the Franck–Condon point, respectively. These differences are smaller than the XMS-CASPT2/cc-pVDZ values (0.58 eV and 0.39 eV), but the "directions" of the energetic differences are the same. These directions differ from the XMS-CASPT2(D)/cc-pVTZ values: 11–12 MECI and 13–14 MECIs are uphill. Still, the 11–12 MECI is more stable than the 13–14 MECI in RPSB6 with XMS-CASPT2(D)/cc-pVTZ. Further benchmarking is warranted as such results are highly dependent on the basis set, shift, and state-averaging schemes. Overall, we have shown that it is possible to use XMCQDPT2 with a resolvent-fitting approximation in a place for XMS-CASPT2 calculations by optimizing the equilibrium and MECI geometries of RPSB6.

**Computational Cost.** Next, let us discuss the computational cost for the resolvent-fitted XMCQDPT2 calculations. We display the wall time associated with the significant computational steps in Table 4 for two test systems: PSB3 with the cc-pV5Z basis set and *p*HBI with the aug-cc-pVTZ basis set. Although the resolvent-fitting approximation is applied, the total wall time for the (12*e*, 11*o*) *p*HBI anion (805 basis functions) is approximately 20 times longer than that for (6*e*, 6*o*) PSB3 (986 basis functions). This is in stark contrast to the internally contracted theories. The parts that are the most sensitive to the active space size in XMS-CASPT2 or QD-NEVPT2 are computations of RDMs and CI derivative evaluations. In resolvent-fitted XMCQDPT2, the energy evaluations, pseudodensities, and CI derivatives are strongly dependent on the number of determinants, as it is an uncontracted theory. For example, the energy, pseudodensity, and CI derivative calculation of *p*HBI required ~30 times more wall time than did PSB3. Still, the size of the CASCI space in *p*HBI is ~500 times larger than that of PSB3. Thus, one can see that the resolvent-fitting approximation is quite efficient in separating the computational efforts that are dependent and not dependent on the active space so that the computational cost does not depend linearly to the size of the CASCI space.

**Table 4.** Wall time for the resolvent-fitted XMCQDPT2 energy and gradient evaluations in seconds.[a] For comparative purpose, wall time for the QD-NEVPT2 and XMS-CASPT2 calculations are also shown.

|  | PSB3 | pHBI anion |
|---|---|---|
| Basis set | cc-pV5Z | aug-cc-pVTZ |
| Active space | (6e, 6o) | (12e, 11o) |
| Model space | 3×3 | 3×3 |
| Number of basis functions | 986 | 805 |
| Denominator, table, and gradient | 6 | 9 |
| 3-index integrals | 5 | 7 |
| Active integrals | 16 | 41 |
| MCQDPT2 energy | 55 | 1832 |
| Pseudodensities and CI derivatives | 533 | 13128 |
| Orbital gradient evaluation | 13 | 16 |
| Z-vector equation | 214 | 833 |
| Derivative integral contraction | 21 | 18 |
| XMCQDPT2 Total time | 863 | 15885 |
| QD-NEVPT2 | 410 | 2195 |
| XMS-CASPT2(D)[b] | 4190 | 11532[c] |
| XMS-CASPT2[b] | 7500 | 26773[c] |

[a] These times are measured using 18 physical cores in an Intel Xeon Gold 6140 processor (2.3 GHz). [b] The so-called "SS-SR" internal contraction scheme was used. [c] Real shift of 0.5 $E_h$ was used to reduce memory requirement.

For comparative purposes, we also performed equivalent calculations with QD-NEVPT2 and XMS-CASPT2 using BAGEL linked with an in-house code for QD-NEVPT2 calculations.[24] In the case of PSB3, QD-NEVPT2 is twice faster than XMCQDPT2, while XMS-CASPT2(D) and XMS-CASPT2 are much more demanding than the XMCQDPT2 calculations. When the active space is larger (*p*HBI anion), the cost of the XMCQDPT2 calculations is comparable to that for the XMS-CASPT2 calculations. An additional advantage of XMCQDPT2 is that it has strict invariance with respect to the perturbation expansion basis (which is only satisfied by employing the "MS-MR" contraction scheme in XMS-CASPT2, which is roughly $N_{\text{state}}$ times more demanding than the "SS-SR" calculations), and ensures the correct behavior near degeneracy. Overall, the current resolvent-fitted XMCQDPT2 algorithm can be an excellent alternative to XMS-CASPT2, particularly with a modest active space size and a large basis set. Of course, we should note that there is some room for improvements for the CASPT2 algorithm in BAGEL, as the fully optimized XMS-CASPT2(D) program should be more efficient than the QD-NEVPT2.

**Inclusion of Many States.** The resolvent-fitting approximation decouples the computational cost proportional to the number of orbitals and determinants from the number of states. Therefore, the inclusion of many states is possible without a significant increase in the computational cost. In contrast, the computational cost is almost proportional to $N_{\text{state}}$ or $N_{\text{state}}^2$ in MS-CASPT2 or QD-NEVPT2 and $N_{\text{state}}$ in the canonical XMCQDPT2. Here, we test the performance of our XMCQDPT2 (which computes perturbation in the uniform FOIS) analytical gradient algorithm when treating a large number of electronic states with (6*e*, 5*o*), (8*e*, 7*o*), (10*e*, 9*o*), and (12*e*, 11*o*) active spaces in the *p*HBI anion. We test the cases of $N_{\text{state}} = 3$ to 10. The relative timings compared to $N_{\text{state}} = 3$ are shown in Figure 2.

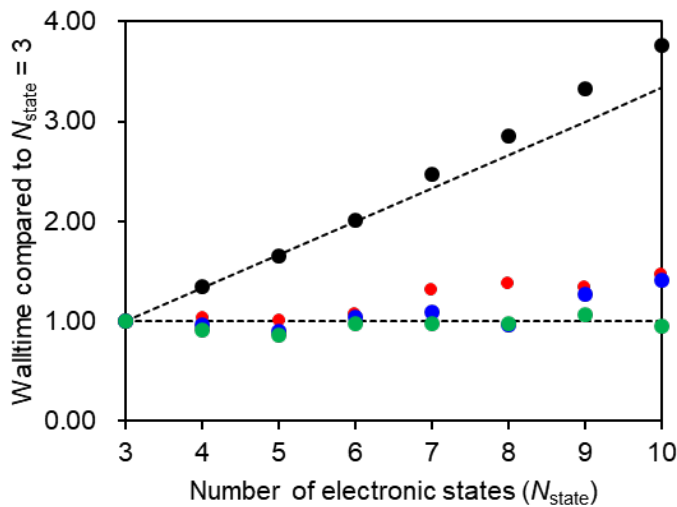

**Figure 2**. Relative wall time of the XMCQDPT2 analytical gradient using various electronic states ($N_{\text{state}}$) compared to $N_{\text{state}} = 3$. The relative wall times for the (12$e$, 11$o$), (10$e$, 9$o$), (8$e$, 7$o$), and (6$e$, 5$o$) active spaces are shown in black, red, blue, and green, respectively. For comparative purposes, the ideal wall times proportional to and constant with respect to $N_{\text{state}}$ are also displayed (dashed lines).

From the (6$e$, 5$o$) to (10$e$, 9$o$) active spaces, the wall times do not significantly increase with respect to the number of states. The most dominant computational effort in the resolvent-fitted XMCQDPT2 is for the three-particle term, which requires

$$N_{\text{grid}}(N_{\text{core}} + N_{\text{vir}})N_{\text{act}}^6 + (N_{\text{fgrid}} + N_{\text{state}})N_{\text{act}}^6 N_{\text{det}} N_{\text{state}} \qquad (40)$$

operations. When $N_{\text{act}}$ and $N_{\text{det}}$ are small, the former term dominates, which is the reason for the small increase in the computational cost with increasing $N_{\text{state}}$. However, there are contributions from the latter term, and when the (10$e$, 9$o$) and (8$e$, 7$o$) active spaces are used, the computational cost slightly increases.

With the (12$e$, 11$o$) active space, the computational cost is almost proportional to $N_{\text{state}}$ [$a$ = 1.08 with the least-squares fit to the curve $(N_{\text{state}}/3)^a$]. In this case, the wall times for both evaluations of the energy and pseudodensity are almost proportional to $N_{\text{state}}$ or slightly higher, as the latter term in Eq. (40) dominates the overall computational cost. If $N_{\text{fgrid}}$ (which is 8 in our implementation) is smaller than or close to $N_{\text{state}}$, the computational cost will be higher than linear.

Overall, the computational cost for the XMCQDPT2 analytical gradient remains almost unchanged with increasing $N_{\text{state}}$ when the active space is small, as the portion proportional to the number of states is minor. However, when the active space has more than ~10 orbitals (or ~100 000 determinants), the cost is proportional to the number of states. Of course, when one employs the XMCQDPT2 analytical gradient for applications such as geometry optimizations or direct dynamics simulations, the stability and continuity of the active space are sensitive to the number of states. Therefore, one should carefully check the numerical stability when one includes a large number of states.

**Parallel Efficiency.** Finally, we comment on the parallel efficiency of the current algorithm with the OpenMP/MPI hybrid approach. OpenMP and MPI are responsible for intranode and internode parallelization, respectively. The zero-particle term is distributed to the MPI process by closed (core) indices, and the evaluations for each active or virtual index are distributed in each core with OpenMP. For one- to three-particle terms, the MPI processes are distributed by two active indices ($N_{\text{act}}^2$). In each process, the one-particle term is distributed by closed (core) or virtual indices in each core, and one-particle virtual-virtual and two- and three-particle terms are distributed by two active indices ($N_{\text{act}}^2$) in each core. For example, the contributions of the second term in the one-particle term in Eq. (7),

$$-\sum_{pqB}\left\langle \alpha\left|\hat{E}_{pq}\right|B\right\rangle c_{B\beta}\sum_{e}\frac{u_{pe}u_{eq}}{\varepsilon_e-\varepsilon_p+\Delta E_{B\beta}} \qquad (41)$$

to the pseudodensities and CI derivatives can be evaluated in parallel as in the pseudocode shown in Fig. 3. The only exception is the virtual-virtual term in the one-particle term

$$-\frac{1}{2}\sum_{pq}\left\langle A\left|\hat{E}_{pq}\right|B\right\rangle \sum_{ia'b'}\frac{(ia'|pb')[2(a'i|b'q)-(a'q|b'i)]}{\varepsilon_{a'}-\varepsilon_i+\varepsilon_{b'}-\varepsilon_q+\Delta E_{B\beta}}, \qquad (42)$$

distributed by $a'$ to exploit the density-fitting strategy introduced at the end of the Theory section.

```
Loop over q,p MPI parallelized
  qp = q + p * N_act
  Loop over e OpenMP threads
    Δ_eq = ε_e - ε_q
    N_peq = u_pe u_eq
    Loop over λ_g
      Δ_eq,g = Δ_eq + λ_g
      D_eq,g = Δ_eq,g / (Δ_eq,g² + τ)
      S⁽¹⁾_pq(λ_g) += - D_eq,g N_peq
      d_e^Fock += (Δ_eq,g² - τ)/(Δ_eq,g² + τ)² P_qp,g N_peq
      d_q^Fock -= (Δ_eq,g² - τ)/(Δ_eq,g² + τ)² P_qp,g N_peq
      d_pe^u -= D_eq,g u_eq P_qp,g'
      d_pe^u -= D_eq,g u_pe P_qp,g'
    End Loop
  End Loop
  … (Evaluate other terms contributing to S⁽¹⁾)
  Loop over α
    Loop over A
      Loop over g∈{1,…,N_fgrid}
        S_pq⁽¹⁾(ΔE_Aα) += W_g(ΔE_Aα) S⁽¹⁾_pq(λ_g)
        G_pq(ΔE_Aα) += (∂W_g(ΔE_Aα)/∂ΔE_Aα) S⁽¹⁾_pq(λ_g)
      End Loop
      Loop over β
        y_Aβ += T_αβ^P <A | E_pq | β> S_pq⁽¹⁾(ΔE_Aα)
        R^β_A,pq += T_αβ^P c_Aα S_pq⁽¹⁾(ΔE_Aα)
        G_Aβ += T_αβ^P c_Aα <A | E_pq | β> G_pq(ΔE_Aα)
      End Loop
    End Loop
  End Loop
End Loop
… (Evaluate all the 2- and 3- particle contributions)
y_Aα += <A | E_qp | B> R^β_B,pq
```

**Figure 3.** Pseudocode for evaluating the contribution of Eq. (41) to the pseudodensities and CI derivatives.

These distributions can deteriorate the parallel efficiency when one uses many MPI processes, but this was not the case for the current tests. We tested the parallel efficiency for (6*e*, 6*o*) PSB3 with cc-pV5Z and (12*e*, 11*o*) *p*HBI with aug-cc-pVTZ. The tests were performed using an Intel Xeon Gold 6140 processor (2.3 GHz, 18 physical cores) connected to an Intel OmniPath 100 Series network node, where a single computer node had two slots for processors (therefore 36 physical cores per node).

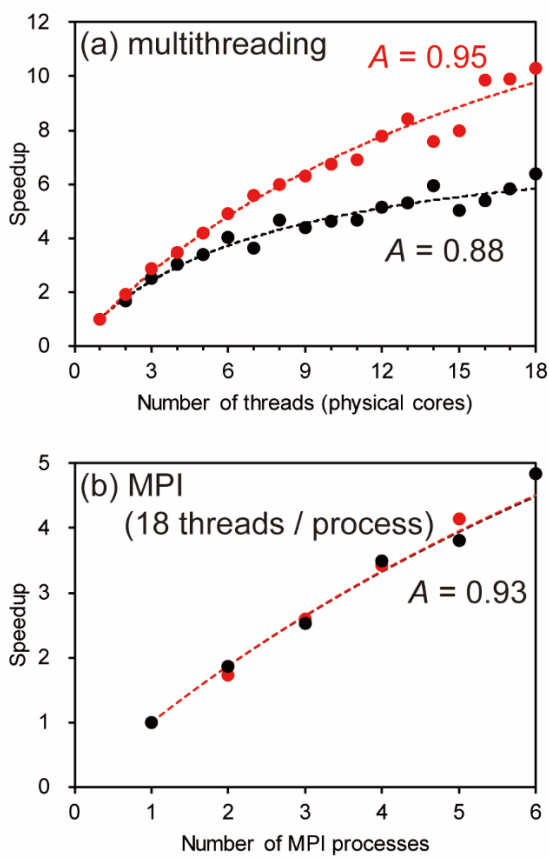

**Figure 4.** Parallel performance of the resolvent-fitted XMCQDPT2 gradient of PSB3 [(6*e*, 6*o*), cc-pV5Z (986 basis functions), black] and *p*HBI [(12*e*, 11*o*), aug-cc-pVTZ (805 basis functions), red]: (a) multithreading performance and (b) internode MPI performance. The data are fitted into Amdahl's law [speedup = $1/(1 - A + A/N)$, where $N$ is the number of threads or processes], and the coefficient $A$ is given.

The current status of multithreading and MPI parallelization is shown in Figure 4, along with Amdahl's law fitting. In the case of multithreading (Figure 4a), Amdahl's law coefficient $A$ was 0.88 and 0.95, respectively, for PSB3 and $p$HBI. There was an anomalous speedup behavior, particularly for PSB3, because we distribute the process with respect to two active orbitals. In PSB3 and $p$HBI, $N_{act}^2$ is 36 and 121, respectively, and the load unbalances are significant when the number of threads is larger than 12. We distribute the jobs statistically, which suggests that the dynamic distributions might improve this behavior. The MPI parallelization (Figure 4b) yielded an A value of 0.93 for PSB3 and $p$HBI. Unlike XMS-CASPT2 or QD-NEVPT2, the parallel efficiency of the algorithm did not significantly degrade when using a large active space. For massively parallel applications, the distributions of the procedure should be redesigned, which will be left for future investigations.

## 4. SUMMARY

In this work, we presented implementations of the analytical gradient of resolvent-fitted XMCQDPT2. This theory has desirable properties such as strict invariance and the lack of dependence on the internally contracted basis. With the resolvent-fitting approximation, the computational cost was comparable to (X)MS-CASPT2 (especially when the basis set and model space are large, and the active space is relatively small), which renders XMCQDPT2 a useful alternative to (X)MS-CASPT2. We demonstrated its applicability for optimizing the geometries of large systems with moderate-sized active spaces. The resolvent-fitting approximation enables geometry optimizations with moderate-sized CASCI spaces up to (12$e$, 12$o$). Furthermore, we parallelized the algorithm using the OpenMP/MPI hybrid strategy. The source codes for the analytical gradient of canonical and resolvent-fitted (X)MCQDPT2 are distributed as a patch on open-source BAGEL (`http://github.com/qsimulate-open/bagel`) version Feb 10 2021 at `http://sites.google.com/view/cbnuqbc/codes` under a GPU-v3 license.

**ACKNOWLEDGMENTS**

This work was supported by a National Research Foundation (NRF) grant funded by the Korean government (MSIT) (Grant 2019R1C1C1003657) and by the POSCO Science Fellowship of POSCO TJ Park Foundation.